\documentclass[%
reprint,
 amsmath,amssymb,
 aps
]{revtex4-2}
\usepackage{hyperref}
\usepackage{xcolor,soul}
\usepackage{mathtools}
\usepackage{verbatim}
\usepackage{graphicx}
\usepackage{dcolumn}
\usepackage{bm}
\usepackage{braket}
\usepackage{enumerate}
\usepackage{enumitem}
\usepackage{simpler-wick}
\usepackage[normalem]{ulem}

\definecolor{darkblue}{rgb}{0,0,0.6}

\newcommand{\bea}{\begin{eqnarray}}
\newcommand{\eea}{\end{eqnarray}}
\newcommand{\be}{\begin{equation}}
\newcommand{\ee}{\end{equation}}

\newcommand{\eq}[1]{\begin{align}#1\end{align}}

\newcommand{\eqsp}[1]{\begin{equation}\begin{split}#1\end{split}\end{equation}}

\newcommand{\lk}{{\mathsf{k}}}
\newcommand{\bfO}{{\mathbf{O}}}
\newcommand{\bfc}{{\mathbf{\chi}}}

\def\a{\alpha}

\newcommand{\overbar}[1]{\mkern 1.5mu\overline{\mkern-1.5mu#1\mkern-1.5mu}\mkern 1.5mu}

\newcommand{\ie}{{i.e.,~}}

\begin{document}

\preprint{APS/123-QED}

\title{Conformal field theories dual to quantum gravity with strongly coupled matter}

\author{Luis Apolo$^{a,b}$, Alexandre Belin$^{c}$, Suzanne Bintanja$^{a}$, \\Alejandra Castro$^{d}$, and Christoph A. Keller$^e$
}
\affiliation{${}^a$Institute for Theoretical Physics, University of Amsterdam, \\ Science Park 904, 1090 GL Amsterdam, The Netherlands
\\
${}^b$Beijing Institute of Mathematical Sciences and Applications, \\ Beijing 101408, China
\\
${}^c$Dipartimento di Fisica, Universit\`a di Milano - Bicocca, \\
I-20126 Milano, Italy
\\
${}^d$Department of Applied Mathematics and Theoretical Physics, University of Cambridge, Cambridge CB3 0WA, United Kingdom
\\
${}^e$Department of Mathematics, University of Arizona, Tuscon, AZ 85721-0089, USA
}

\date{\today}

\begin{abstract}

A holographic conformal field theory is dual to semi-classical general relativity in Anti-de Sitter space coupled to matter fields. If the CFT factorizes in the large-$N$ limit, then all couplings in its dual are suppressed by the Planck scale, making the matter fields weakly interacting. We propose a mechanism to produce CFTs whose dual matter fields couple weakly to gravity, but interact strongly with each other. We achieve this by turning on exactly marginal multi-trace deformations, and quantify the effect using conformal perturbation theory.

\end{abstract}

\maketitle


\section{\label{sec:intro}Introduction}

What is the space of consistent theories of quantum gravity? How can the Standard Model be consistently coupled to gravity? Answering these questions remains an important challenge in understanding the laws of physics in our universe. For theories of gravity with a negative cosmological constant, i.e.,~in Anti-de Sitter (AdS) space, much progress towards answering these questions has been made through the AdS/CFT correspondence \cite{Maldacena:1997re}. In AdS, the rules for obtaining a consistent theory at the quantum level are clear. Standard observables of quantum gravity in AdS$_{d+1}$ are conformal correlation functions of local operators inserted at the asymptotic boundary. These correlators must satisfy the axioms of a conformal field theory in $d$ dimensions (CFT$_d$): unitarity, causality, and crossing symmetry.

However, not all CFTs will be dual to semi-classical gravity minimally coupled to matter fields. CFTs that have gravity duals of this type are known as \textit{holographic} CFTs, and they must satisfy very stringent constraints: they must possess a large number of degrees of freedom such that the stress tensor 2-point function $\braket{TT}\sim N$ for some large $N$; they should have a sparse number of operators with low scaling dimension \cite{Hartman:2014oaa,Belin:2016yll,Mefford:2017oxy}; and they should have a large gap in the scaling dimension of higher spin operators \cite{Heemskerk:2009pn,Afkhami-Jeddi:2016ntf,Meltzer:2017rtf,Belin:2019mnx,Kologlu:2019bco,Caron-Huot:2021enk}.

Holographic CFTs also have a set of operators whose correlation functions factorize in the large-$N$ limit \cite{ElShowk:2011ag}. However, it is a common misconception that all operators must factorize. This misconception may come from the fact that the best known holographic CFTs are large-$N$ gauge theories in the 't Hooft (or planar) limit; there, all operators indeed factorize in the large-$N$ limit. 

To explain why this does not necessarily have to be the case,
consider a holographic CFT  with a stress tensor and an additional single-trace scalar operator $\bfO$ that satisfies large-$N$ factorization. In the gravity dual, the effective Euclidean action is of the form
\eq{ \label{actionweak}
&S_{\textrm{bulk}}=-\frac{1}{16\pi G_N} \int d^{d+1}x \sqrt{g} \left( R+ \frac{d(d-1)}{\ell_{\textrm{AdS}}^2} \right) \\
&+\int d^{d+1}x \sqrt{g} \left(\frac{1}{2}\partial_\mu \phi \partial^\mu \phi + \tilde{\mathsf{g}}_3 \frac{G_N^{1/2}}{\ell_{\textrm{AdS}}^{2}} \phi^3 + \tilde{\mathsf{g}}_4 \frac{G_N}{ \ell_{\text{AdS}}^2} \phi^4 \right) \notag,
}
where $\phi$ is the bulk field dual to $\bfO$. In the semi-classical limit where $\ell_{\text{AdS}}^{d-1}/G_N\sim N\to\infty$ and for $\tilde{\mathsf{g}}_{3,4}\sim\mathcal{O}(1)$, this action describes a theory where all interactions happen at the Planck scale.  This is manifest for gravitational interactions, but is also true for interactions of the scalar field due to the explicit $G_N$ dependence in front of its couplings. As a result, such a holographic CFT can never describe gravitational theories where the matter sector is strongly interacting, such as the Standard Model.

In this paper, we are interested in constructing holographic CFTs whose gravitational dual contains a strongly interacting matter sector. As explained above, this means that the dual CFT cannot satisfy large-$N$ factorization: correlation functions of the stress tensor will factorize, but those of the operator $\bfO$ dual to $\phi$ will not. For example, we want to construct theories where the three-point function of the operator $\bfO$ scales as 
\eq{
\left\langle \bfO \bfO \bfO\right\rangle \sim \mathcal{O}(N^0) \,,\label{eq:3ptstrong}
}
instead of scaling like $N^{-1/2}$ as would be the case in a theory that obeys large-$N$ factorization. On the other hand, the stress-tensor sector should still factorize at large $N$ as we only want the matter to be strongly interacting, and gravitational interactions as well as their coupling to the matter sector should still be weak. To the best of our knowledge, not a single theory of this type has been explicitly constructed \footnote{One can obtain strongly coupled matter in AdS by adding branes in the bulk on which the matter is strongly coupled, or by considering RG flows which can flow to strongly coupled matter in the infrared, but such setups will either involve a relevant deformation of a CFT, or adding interfaces.}. 

To obtain such theories, we break large-$N$ factorization by deforming CFTs by exactly marginal multi-trace operators. These deformations were considered in \cite{Aharony:2001pa,Aharony:2001dp} albeit in a regime where large-$N$ factorization is preserved. We deform the CFT action by
\be\label{eq:def}
S_{\textrm{CFT}}\mapsto S_{\textrm{CFT}} +\lambda N^{\frac{\beta}{2}} \int d^{d}x \  \bfO^\lk(x) \,.
\ee
Here $\bfO^\lk$ is an exactly marginal $\lk$-trace operator such that $\Delta_{\bfO^\lk}=d$ independently of $\lambda$.
Crucially, we can scale the deformation in an $N$-dependent way ($\lambda$ will always be independent of $N$  in our conventions). The precise form of this scaling determines the large-$N$ properties of the deformed theory.
To preserve large-$N$ factorization, we must have $\beta\leq 2-\lk$ \cite{Aharony:2001pa,Belin:2020nmp}. From the bulk perspective, the upper bound $\beta=2-\lk$ can induce shifts in the bulk couplings  of the form $\tilde{\mathsf{g}}_{3,4}\mapsto\tilde{\mathsf{g}}_{3,4}+\lambda$.

In this paper, we take a different $N$ scaling for the deformation, namely one where $\beta=0$. The deformed CFT no longer factorizes, and has a bulk dual where the matter fields are coupled at the AdS scale rather than at the Planck scale. Our deformation can continuously interpolate between holographic CFTs that factorize and theories that admit a strongly coupled matter sector.

The mechanism is quite general and can be applied to any theory that contains an exactly marginal multi-trace operator. As an example of such theories, we will consider the two-dimensional CFTs with $\mathcal{N}=2$ supersymmetry discussed in \cite{Belin:2020nmp}.  

\section{Large-\textit{N} CFTs}\label{sec:largen}

Our objective is to realize a CFT that has a well-defined large-$N$ limit, but does not obey large-$N$ factorization. To achieve this, we start with a theory that does satisfy large-$N$ factorization and deform it by a multi-trace deformation that breaks the factorization property in a controlled manner. In particular, we will be able to track the effect of the deformation order by order in the $1/N$ expansion using large-$N$ factorization of the undeformed theory. We start by reviewing salient features of large-$N$ CFTs and large-$N$ factorization (see \cite{ElShowk:2011ag} for details).

We define a CFT with a large-$N$ limit as follows: it contains a dimensionless parameter $N$  controlling the stress tensor 2-point function as $\langle TT\rangle\sim N$, which is taken to be large. Its operators fall into two categories \footnote{There can be additional intermediate regimes; see, e.g., \cite{Hartman:2014oaa}}: {\it light} operators with scaling dimension $\Delta\sim N^0$, whose degeneracy is independent of $N$; and {\it heavy} operators, whose dimension grows with $N$ \footnote{The heavy operators, at least those whose dimensions scale linearly with $N$, are often referred to as black hole microstates.}. In the following, we will always normalize operators such that they have unit 2-point functions. Moreover, all correlation functions of light operators must converge as $N\rightarrow\infty$. 

We say a theory factorizes in the large-$N$ limit if the light spectrum satisfies the following rules: there is a set of \textit{single-trace} operators that behave as generalized free fields as $N\to\infty$. More precisely, their connected correlation functions scale as
\eq{\label{eq:thooft}
\left\langle \bfO_1(x_1)\cdots \bfO_n(x_n)\right\rangle_c\sim N^{-(n-2)/2}\,.
}
The leading order contribution to a correlation function of single-trace operators thus comes from the disconnected part where as many operators as possible are Wick contracted.

There are also sectors of \textit{multi-trace} operators $\bfO^\lk$ of trace $\lk$, defined by the fact that there exist single-trace operators $\bfO_1,\dots,\bfO_\lk$, such that
\eq{\label{eq:mt}
\left\langle \bfO^\lk(x)\bfO_1(x_1)\cdots \bfO_\lk(x_\lk)\right\rangle_c\sim N^0\,.
}
$\lk$-trace operators are thus normal ordered products of $\lk$ single-trace operators $\bfO^\lk= \ :\!\bfO_1\cdots\bfO_\lk\!:$, possibly including derivatives. A correlation function of  multi-trace operators can be computed as a sum over products of connected correlation functions of their component single-trace operators. From (\ref{eq:thooft}), each single trace correlator contributes a factor of $N^{-1/2}$ for each operator in excess of two, so that the leading contribution to the correlation function is given by the terms with minimal excess operators. In particular, in many cases there are $\mathcal{O}(N^0)$ connected contributions achieved solely from Wick contractions of the single-trace components. This behavior is very different than the scaling of single-trace correlators, and will be crucial to break factorization. 

If single and multi-trace operators fully span the Hilbert space of light states, then the CFT satisfies large-$N$ factorization. Examples of such CFTs are large-$N$ vector models and large-$N$ gauge theories with adjoint matter, and, in two dimensions, symmetric product orbifolds and certain permutation orbifolds \cite{Lunin:2000yv,Pakman:2009zz,Belin:2015hwa,Gemunden:2022xux}.

Now we want to deform such large-$N$ CFTs by an exactly marginal operator. A general deformation is of the form
\eq{
S_{\textrm{CFT}}\mapsto S_{\textrm{CFT}} + \lambda_\lk N^{\beta/2}\int d^{d}x \  \bfO^\lk(x) \,, \label{eq:defdef}
}
where $\lambda_\lk\sim N^0$ and $\bfO^\lk$ is a normalized exactly marginal $\lk$-trace operator. The interplay between the parameters $\beta$ and $\lk$ leads to various effects in the deformed theory:
\begin{description}[leftmargin=0.0cm]

\item[Existence of the large-$\bm{N}$ limit] All quantities in the perturbed CFT need to converge as $N \to \infty$. This is automatically the case if we choose a non-positive $\beta$, since all correlators of the perturbed theory are given by higher-point functions in the undeformed theory, which we know converge. However, sometimes it is also possible to choose $\beta$ positive: for instance, $\beta=1$ for single-trace deformations still leads to a convergent large-$N$ limit, because the effect of the deformation comes from the connected components of correlation functions only, which scale with negative powers of $N$.

\item[Lifting unprotected operators] 
Deformations with $\beta\leq 0$ give $\mathcal O(N^{-1})$ effects to the anomalous dimensions of most operators \footnote{For double-trace deformations with $\beta=0$, some operators (those who make up the double-trace deformation) can acquire $\mathcal{O}(N^0)$ anomalous dimensions. However, all other single-trace operators only acquire anomalous dimensions at $\mathcal{O}(N^{-1})$}. To obtain a lifting that affects most operators but still leads to a convergent large-$N$ limit, one must necessarily use single-trace deformations (with $\beta=1$). In particular, this shows that to deform away from a free point of the conformal manifold, where there are infinite towers of conserved currents, and reach a holographic point, one must use single-trace deformations. This has been explicitly checked for two-dimensional CFTs in \cite{Avery:2010er,Gaberdiel:2015uca,Keller:2019yrr,Guo:2020gxm, Apolo:2022fya, Benjamin:2022jin}, and could also be studied in the gravitational dual, at least close to the tensionless string theory point, see \cite{Gaberdiel:2014cha,Giribet:2018ada,Eberhardt:2018ouy,Eberhardt:2019ywk}.

\item[Large-$\bm{N}$ factorization] 
If $\beta\leq 2-\lk$, then the deformed theory still complies with large-$N$ factorization (see App. D of \cite{Belin:2020nmp}). To break factorization, we therefore want to consider a $\lk$-trace deformation with $\beta_\lk$ satisfying $ 2-\lk < \beta_\lk\leq0$. In particular, for deformations that are at least triple-trace ($\lk>2$), we can set $\beta_\lk=0$ and break large-$N$ factorization while preserving the convergence of the large-$N$ limit. 
\end{description}

We thus find that there is an important difference between single and multi-trace deformations. Single-trace deformations can lift the unprotected operators, for example breaking higher spin symmetries; they are thus needed to reach a holographic point in moduli space. However, they always preserve large-$N$ factorization. Multi-trace deformations on the other hand, can break factorization, allowing us to explore strong interactions in the bulk, but they cannot lift the majority of operators. To obtain a holographic CFT with strongly coupled matter, a two-step process is needed: first, lift unprotected operators using a large single-trace deformation, preserving factorization. Second, starting from that point on the conformal manifold, break factorization using a multi-trace deformation. 

\section{\label{sec:cpt} Breaking large-\textit{N} factorization}

In this section, we will perform an explicit computation to show how multi-trace marginal deformations break large-$N$ factorization in two-dimensional CFTs with $\mathcal N = 2$ supersymmetry. We use a triple-trace deformation, which will affect OPE coefficients of single-trace operators at $\mathcal{O}(N^0)$ \footnote{One may wonder why the case of $\lk=2$ is not considered; it is more subtle. If the double-trace operator is made out of two protected single-trace components, which is likely always the case, then it turns out to have no effect at $\mathcal O(N^0)$. Its only potential effect at that order is on the 2-point function of its constituents, but these are protected and hence their 2-point function is not modified. Because of this, one may try to set $\beta=1$ for a double-trace deformation, but we expect higher order corrections in $\lambda_2$ to diverge in $N$.}. 

To ensure that the deformation is exactly marginal, \ie that its dimension does not receive corrections, we use supersymmetry. In a two-dimensional CFT with $\mathcal{N}=2$ supersymmetry, we can obtain exactly marginal operators from the superpartner $G^-_{-1/2} \overbar G^-_{ -1/2}\bfO^\lk$ of a chiral primary $\bfO^\lk$  of weight $1/2$ and $U(1)$-charge $1$. Exactly marginal multi-trace deformations of this type appear for instance in symmetric product orbifolds of $\mathcal N=2$  minimal models \cite{Belin:2020nmp}. In our case, we consider a triple-trace deformation 
\eq{
\bfO^3(z) \coloneqq \frac{1}{{4\sqrt{3}}} \big( G^-_{\scriptscriptstyle -1/2} \overbar G^-_{\scriptscriptstyle -1/2}  :\! \bfO \bfO \bfO \!:\!(z) + \text{h.\,c.} \big) \label{eq:tripletrace}\ ,
}
with $\bfO(z)$ a single-trace chiral primary of conformal weight $1/6$ and $U(1)$ charge $1/3$ \footnote{This modulus appears for example in the symmetric orbifold of the $A$-series $\mathcal{N}=2$ minimal models $A_{k+1}$, with $k=1\ \text{mod} \ 3$, and $k\ge4$. It also appears in the symmetric orbifold of the $D$-series $\mathcal{N}=2$ minimal models $D_{k/2+2}$ whenever $k=4\ \text{mod}\ 6$. There are also many other exactly marginal triple-trace  and higher-trace operators for different choices of the constituent operators \cite{Belin:2020nmp}}.

The triple-trace deformation \eqref{eq:tripletrace} changes the 3-point function of single trace operators $\bfO_i$ as 
\eqsp{
& \nabla_{\lambda_3}   C_{\bfO_i\bfO_j\bfO_k}\coloneqq \\
& \hspace{55pt} \int \! d^2x  \langle \bfO^3(x) \bfO_i(\infty) \bfO_j(1)  \bfO_k(0) \rangle_c \Big|_{\textrm{reg}}~, \label{eq:dOPE}
}
where $C_{\bfO_i\bfO_j\bfO_k}$ is the OPE coefficient between the operators $ \bfO_i$, $\bfO_j$, and $\bfO_k$, $\nabla_{\lambda_3}$ denotes the covariant derivative on the conformal manifold, and $\textrm{reg}$ indicates that we have regularized the integral as discussed in App.~\ref{sec:integral}. Note that the covariant derivative takes into account the fact that the conformal manifold is generically curved, although this cannot be seen at leading order in perturbation theory~\cite{Ranganathan:1993vj}.

Specifically, we consider the OPE coefficient $C_{\bfO\bfO\bfc}$ between two $\bfO(z)$ operators and
\eq{
\bfc(z) \coloneqq \frac{3}{2}G^-_{ -1/2} \overbar G^-_{ -1/2}\bfO(z)\,.
}
The reason for this is that, at $\lambda_3=0$ and in the large-$N$ limit, the triple-trace deformation  can only affect OPE coefficients of the constituents of the modulus. This follows from the fact that only fully Wick-contracted terms contribute to correlation functions at order $\mathcal{O}(N^0)$. 

We now show that \eqref{eq:dOPE} for  $C_{\bfO\bfO\bfc}$ is nonvanishing at the point $\lambda_3=0$. 
Note that, unlike for $\mathcal N = 4$ SCFTs  \cite{deBoer:2008ss,Baggio:2012rr}, there is no nonrenormalization theorem that forces \eqref{eq:dOPE} to vanish here. We thus need to evaluate the following connected 4-point function 
\eq{
\mathcal I(x) \coloneqq \langle \bfO^3(x) \bfO(\infty) \bfO(1)\bfc(0) \rangle_c\, .\label{eq:Idef}
} 
As mentioned above, this 4-point function is determined by Wick contractions of pairs of $\bfO(z)$ and pairs of $\bfc(z)$ fields such that
\eq{
\mathcal I(x) &= \frac{1}{2\sqrt{3}}\, \big\langle \!:\!\bfc^\dagger \bfO^\dagger\bfO^\dagger\!\!:\!(x) \bfO(\infty) \bfO(1)\bfc(0) \big\rangle_c \\
&=\frac{1}{2\sqrt{3}}\, \big\langle \wick{\!:\!\c1\bfc^\dagger \c2\bfO^\dagger \c3\bfO^\dagger\!\!:\!(x) \c3 \bfO(\infty) \c2\bfO(1) \c1\bfc(0)} \big\rangle + \dots \\
&=\frac{1}{\sqrt{3}}\frac{1}{|x|^{8/3}|x-1|^{2/3}} + \mathcal O (N^{-1}) \,, \label{eq:4ptresult}
}
where in the first line we only kept the nonvanishing contributions of $\bfO^3(x)$, and in the second line we omitted terms suppressed by factors of $N^{-1}$ as well as another set of Wick contractions that accounts for the extra factor of 2 in the third line \footnote{Note that the large-$N$ limit, \eqref{eq:4ptresult} is fully determined by the conformal weight of the chiral primary $\bfO(z)$. Supersymmetry guarantees that the conformal weight of this field is protected on the conformal manifold. Consequently, \eqref{eq:4ptresult} is valid (nonperturbatively) under any other exactly marginal deformations that preserve large-$N$ factorization.}. 

Therefore, to leading order in perturbation theory we have
\eqsp{
&\nabla_{\lambda_3} C_{\bfO\bfO\bfc}\big|_{\lambda_3 = 0} = \\ 
& \hspace{35pt} \frac{1}{\sqrt{3}}\int d^2x \, \frac{1}{|x|^{8/3}|x-1|^{2/3}} \bigg|_{\text{reg}} + \mathcal O(N^{-1}) \,. \label{eq:nablaC0}
}
This integral can be regularized to yield a finite result, as discussed in App.~\ref{sec:integral}, and it evaluates to
\eq{
\nabla_{\lambda_3} C_{\bfO\bfO\bfc}\big|_{\lambda_3 = 0} =-\frac{8\pi^4}{\Gamma(1/3)^6} + \mathcal O(N^{-1})\,. \label{eq:nablaC}
}
As explained there, there is no operator mixing between BPS states at first order in the large-$N$ limit. The shift in the 3-point function is thus a genuine effect, and not the result of a mixing between single and multi-trace operators.

The upshot is the following: the 3-point function $C_{\bfO\bfO\bfc}$, which in the large-$N$ limit vanishes at $\lambda_3=0$ due to factorization, receives a non-vanishing correction. Turning on the deformation thus breaks large-$N$ factorization for the operators $\chi$ and $\bfO$. 

The implications for the AdS theory are interesting. From \eqref{eq:nablaC} we would generally infer that there is a 3-point interaction among the bulk fields $\phi$ and $\psi$ (dual to $\chi$) which is controlled by the AdS scale, and independent of $G_N$.  This can be viewed as a signal of strong coupling among matter fields, albeit the calculation presented above is in conformal perturbation theory and the bulk theory would only be strongly coupled at finite $\lambda$.  It is therefore tempting to write an effective action in AdS of the form
\eqsp{ \label{actionstrong}
S_{\textrm{matter}}&=\int d^{d+1}x \sqrt{g} \Big(\frac{1}{2}\partial_\mu \phi \partial^\mu \phi +\frac{1}{2}\partial_\mu \psi \partial^\mu \psi  \\
&\hspace{80pt}+ \mathsf{g}_3 \ell_{\textrm{AdS}}^{\frac{d-5}{2}} \phi^2 \psi + \cdots  \Big) \,, 
}
where $\mathsf{g}_{3}\sim\mathcal{O}(G_N^0)$. This action would indeed describe  a strongly interacting matter sector at the AdS scale.

However, the effect we observe in \eqref{eq:nablaC} cannot in fact correspond to a  local interaction in AdS: since $\Delta_\bfO+\Delta_\bfO+\Delta_\chi=d$, the interaction will most likely be captured by a total derivative or a boundary term instead \footnote{For $\Delta_\bfO+\Delta_\bfO+\Delta_\chi=d$ there is a divergence in Witten diagrams, analogous to those in extremal correlators \cite{DHoker:1999jke}. We thank Shota Komatsu for his remarks on this issue.}, but its precise bulk interpretation needs further study. We discuss this in more detail below.

\section{Discussion}

In this paper, we have presented a mechanism to obtain holographic CFTs whose matter sector is strongly coupled. The mechanism works by deforming a holographic CFT by an exactly marginal multi-trace deformation. The deformation preserves a convergent large-$N$ limit, but destroys large-$N$ factorization; it allows to continuously interpolate between a holographic CFT with large-$N$ factorization and a theory whose bulk matter sector is strongly coupled. We conclude with some open questions and future directions.

The most important question remains what the proper AdS$_{d+1}$ interpretation of our result is. As we have discussed in \eqref{actionstrong}, a cubic coupling cannot be written in the effective AdS$_{d+1}$ Lagrangian as it leads to an IR-divergence in the Witten diagram. Note that this phenomenon is already true in the undeformed CFT that complies with large-$N$ factorization, namely the action \eqref{actionweak} is not the right effective theory for the bulk fields. It is likely that local bulk interactions are replaced by boundary terms (a similar effect was found for $\Delta=1$ operators in ABJM \cite{Freedman:2016yue}), and for our deformations, the bulk matter is made strongly coupled by a strong boundary interaction rather than a strong bulk coupling. It remains to be understood whether there is any physical significance in the difference between the two types of interactions at the level of CFT correlation functions, which are the physical observables. We hope to return to this question in the future.

Multi-trace deformations with $\lk>3$ should correspond to changing higher-point couplings in the bulk. It would be interesting to study this further. Note that in the theories of \cite{Belin:2020nmp}, exactly marginal operators of arbitrary trace were explicitly identified; other examples include the theories studied in \cite{Benjamin:2022jin}.

On a related front, it would also be interesting to study the effects of the triple-trace deformation to higher orders in conformal perturbation theory. At second order, the deformation we considered in this paper would affect 4-point functions. This is natural since two operators $\bfO$ can fuse into the operator $\chi$, so this will induce a change in the four-point function $\braket{\bfO\bfO\bfO\bfO}$ at second order in $\lambda$. It is possible that the triple-trace deformation also introduces a quartic coupling in the bulk effective Lagrangian. This could be investigated by directly computing the deformed 4-point function to second order in conformal perturbation theory, and we hope to return to this question in the future.

To obtain strongly coupled bulk matter, we must go beyond conformal perturbation theory and make the coupling finite. It appears challenging to tackle this problem head-on. After all, by construction we are studying a strongly coupled field theory in AdS. Nevertheless, the goal of this paper was  to introduce the mechanism that makes this situation possible. It is worthwhile to mention what we see as the only potential loophole in our construction. It is possible that the conformal manifold is actually periodic in the direction of the triple-trace deformation. In fact, conjectures in the swampland program state that most directions on a conformal manifold should be compact (see \cite{Perlmutter:2020buo} in the context of AdS/CFT). From our point of view, the crucial question is to understand the compactification radius.  If it is $\mathcal{O}(1)$ in units of $\lambda_3$ with $\beta=0$ as given in \eqref{eq:defdef}, the mechanism we propose holds. If it scales as $N^{-m}$ for some positive $m$, our construction would break down. We do not expect this to be the case, because the exactly marginal deformation exists as long as $N>3$, and therefore one would expect its compactification radius to be typically $\mathcal{O}(1)$, if it is compact at all. It would be very interesting to study this question further, as very little is known about global properties of conformal manifolds.

It would also be interesting to find more top-down constructions where multi-trace exactly marginal operators exist. An exactly marginal double-trace operator in the Klebanov-Witten CFT which has $\mathcal{N}=1$ in $d=4$ was discussed in \cite{Aharony:2001pa}. To the best of our knowledge, the exactly marginal operators with $\lk>2$ described in \cite{Belin:2020nmp} are currently the only known examples (along with symmetric orbifolds of tensor products of minimal models \cite{Benjamin:2022jin}). It is however likely that many more can be found. Note that for a triple-trace exactly marginal operator to exist, we must have $d<6$, and the restriction becomes stronger as we increase the number of traces. For six-trace operators and beyond, we must be in $d=2$. It is also important to mention that the theories in \cite{Belin:2020nmp} have not yet been shown to be holographic, even though there is evidence both from BPS quantities and in conformal perturbation theory \cite{Apolo:2022fya}. It would thus be particularly interesting to find triple-trace marginal operators in known holographic theories. 

It is also worth discussing the fate of bulk locality. In \cite{Aharony:2001pa, Aharony:2005sh}, possible issues with bulk locality were raised. In the Klebanov-Witten theory, this can be seen from the fact that the multi-trace operator is built out of particular Kaluza-Klein modes on $T^{1,1}$. This would render the deformation non-local on $T^{1,1}$ and in \cite{Aharony:2005sh}, it was argued that the theory should also be non-local on the AdS space.  In our construction, we do not see an explicit manifestation of non-locality;  our interactions are likely captured by boundary terms, but this does not necessarily imply that there are non-local effects. Non-locality would appear in the bulk effective theory of the fields $\phi$ and $\psi$ (the bulk field dual to $\bfc$) through higher derivative couplings. In our case, while some quartic couplings may be generated, only a finite number of higher derivative terms are allowed to appear. This follows from the fact that there are only exchanges of fields of finite spin, see for example \cite{Camanho:2014apa}. Therefore, we do not seem to find any issue with locality, at least in conformal perturbation theory. It would nevertheless  be interesting to study this question further. 

Finally, we would like to comment on the possibility that making the matter strongly coupled could in turn make the gravitational sector (i.e.~the stress tensor sector) strongly coupled. In the $d=2$ case, this never happens because the strength of gravitational interactions is universally determined by the central charge of the CFT, which is unchanged by the deformation. For $d>2$, the situation is less clear. It is an open problem to show that gravitational interactions are weak in the dual of any CFT with large $N$. In any case, for the mechanism described in this letter, we are assured that gravitational interactions remain weak as long as the stress tensor is not used as a component of the multi-trace deformation.
\vspace{-1mm}
\begin{acknowledgments}
We thank Ofer Aharony, Nikolay Bobev, Shota Komatsu, Juan Maldacena, Kyriakos Papadodimas, Jo\~{a}o Penedones, Eric Perlmutter, Mukund Rangamani, and Sasha Zhiboedov for interesting discussions, and collaborations on related topics. LA thanks the Asia Pacific Center for Theoretical Physics (APCTP) for hospitality during the focus program ``Integrability, Duality and Related Topics", as well as the Korea Institute for Advanced Study (KIAS) for hospitality during the ``East Asia Joint Workshop on Fields and Strings 2022'', where part of this work was completed. The work of LA is supported by the Dutch Research Council (NWO) through the Scanning New Horizons programme (16SNH02). SB thanks DAMTP for hospitality. The work of SB is supported by the Delta ITP consortium, a program of the Netherlands Organisation for Scientific Research (NWO) that is funded by the Dutch Ministry of Education, Culture and Science (OCW). The work of AC has been partially supported by STFC consolidated grant ST/T000694/1. The work of CAK is supported in part by the Simons Foundation
Grant No. 629215 and by NSF Grant 2111748.

\end{acknowledgments}

\appendix

\section{Regularization of $\mathbf{C_{\bfO\bfO\bfc}}$}\label{sec:integral}
In this appendix we evaluate the integral
\eq{
I = \int d^2x \,\frac{1}{|x|^{8/3}|x-1|^{2/3}}\,, \label{eq:theintegral}
}
which determines the covariant derivative of the OPE coefficient $C_{\bfO\bfO\bfc}$ on the conformal manifold \eqref{eq:nablaC0}. To regularize the integral, we use a hard-sphere cutoff with minimal subtraction scheme \cite{Ranganathan:1993vj}. That is, we cut out $\epsilon$-discs around divergent points, subtract only terms divergent in $\epsilon$, and send $\epsilon \to 0$. We note that in this scheme the first derivative of the Zamolodchikov metric vanishes at $\lambda_3=0$ \cite{Kutasov:1988xb,Friedan:2012hi}. Similarly, at first order two operators that do not get lifted do not mix, because in that case their 3-point function with the modulus vanishes, and we do not add or subtract any constant terms. 

The statements above are general properties for conformal perturbation theory in any CFT$_2$. In our case, the situation is even simpler due to the large-$N$ limit. Mixing with higher trace operators or with other single-trace operators that share the same quantum numbers involves relative factors of $\mathcal O(N^{-\alpha})$ for some $\a > 0$, which follows from large-$N$ factorization at $\lambda_3 = 0$. Therefore, mixing between operators is absent in any regularization scheme to leading order in the large-$N$ limit. Subleading $\mathcal O(N^{-1})$ corrections to \eqref{eq:nablaC} coming from mixing are possible, and expected \cite{Arutyunov:1999en,Arutyunov:2000ima}.

We can evaluate the integral using radial coordinates $x = r e^{i \theta}$ by performing first the integration over $\theta$
\eq{
I &= \int_0^\infty dr \ \int_0^{2\pi}d\theta \ \frac{1}{r^{5/3}\left(1+r^2-2r\cos(\theta)\right)^{1/3}}\, \notag\\
&= 2 \pi \int_0^\infty dr \ \frac{  \, _2F_1\left(\frac{1}{3},\frac{1}{2};1;\frac{4}{\alpha + 2}\right)}{(\alpha + 2)^{1/3} r^2}\,, \quad \alpha \coloneqq r+\frac{1}{r}\,. \label{eq:3}
}
By exploiting the invariance of $\alpha$ under $r\mapsto\frac{1}{r}$ and using properties of the hypergeometric functions we find that \eqref{eq:3} can be simplified to
\eq{
I &= 2\pi  \int_0^1 dr \ \frac{\left(r^2+1\right) \, _2F_1\left(\frac{1}{3},\frac{1}{3};1;r^2\right)}{r^{5/3}}\,.\label{eq:9}
}
The integral \eqref{eq:9} can be evaluated analytically but diverges at $r = 0$. Following our regularization scheme, we introduce a cutoff $\epsilon$ to obtain
\eq{
I&=\frac{3 \pi }{\epsilon^{2/3}}-\frac{8 \sqrt{3} \pi ^4}{\Gamma (1/3)^6} +\mathcal{O}\big(\epsilon^{4/3}\big)\,.
}
We then use minimal substraction to eliminate the first term. As a result, the renormalized value of the integral \eqref{eq:theintegral} is finite and given by
\eq{
I_{\text{reg}} &= -\frac{8 \sqrt{3} \pi ^4}{\Gamma (1/3)^6}.
}

\bibliography{ref.bib}

\end{document}